# A giant outburst two years before the core-collapse of a massive star

A. Pastorello[1], S.J. Smartt[1], S. Mattila[1], J.J. Eldridge[1], D. Young[1], K. Itagaki[2], H. Yamaoka[3], H. Navasardyan[4], S. Valenti[5,6], F. Patat[5], I. Agnoletto[4,7], T. Augusteijn[8], S. Benetti[4], E. Cappellaro[4], T. Boles[9], J.-M. Bonnet-Bidaud[10], M.T. Botticella[11], F. Bufano[4,7], C. Cao[12], J. Deng[12,13], M. Dennefeld[14], N. Elias-Rosa[4,15], A. Harutyunyan[4,7], F.P. Keenan[1], T. Iijima[16], V. Lorenzi[17], P.A. Mazzali[18,19], X. Meng[12], S. Nakano[20], T.B. Nielsen[8], J.V. Smoker[1], V. Stanishev[21], M. Turatto[4], D. Xu[12], L. Zampieri[4]

1. Astrophysics Research Centre, School of Mathematics and Physics, Queen's University Belfast, Belfast BT7 1NN, United Kingdom

2. Itagaki Astronomical Observatory, Teppo-cho, Yamagata, Japan

3. Department of Physics, Kyushu University, Fukuoka 810-8560, Japan.

4. INAF Osservatorio Astronomico di Padova, Vicolo dell' Osservatorio 5, I-35122 Padova, Italy

5. European Southern Observatory (ESO), Karl-Schwarzschild-Str. 2, D-85748 Garching bei München, Germany

6. Dipartimento di Fisica, Università di Ferrara, via del Paradiso 12, I-44100 Ferrara, Italy

7. Dipartimento di Astronomia, Università di Padova, Vicolo dell'Osservatorio 2, I-35122 Padova, Italy

8. Nordic Optical Telescope, Apartado 474, E-38700 Santa Cruz de la Palma, Tenerife, Spain

*9. Coddenham Astronomical Observatory, Suffolk, United Kingdom*

*10. Service d'Astrophysique, DSM/DAPNIA/SAp, CE Saclay, F-91191 Gif-sur-Yvette Cedex, France*

*11. Dipartimento di Scienze della Comunicazione, Università degli Studi di Teramo, Viale Crucioli 122, I-64100 Teramo, Italy*

*12. National Astronomical Observatories, Chinese Academy of Sciences, 20A Datun Road, Chaoyang District, Beijing 100012, China*

*13. Centre de Physique des Particules de Marseille, CNRS-IN2P3 and University Aix Marseille II, Case 907, 13288 Marseille Cedex 9, France*

*14. Institut d'Astrophysique de Paris, CNRS, and Université P. et M. Curie, 98bis Bd Arago, F-75014 Paris, France*

*15. Universidad de La Laguna, Av. Astrofísico Francisco Sánchez s/n, E-38206 La Laguna, Tenerife, Spain*

*16. INAF Osservatorio Astronomico di Padova, Sezione di Asiago, Via dell'Osservatorio 8, I-36012 Asiago (Vicenza), Italy*

*17. Fundación Galileo Galilei-INAF, Telescopio Nazionale Galileo, E-38700 Santa Cruz de la Palma, Tenerife, Spain*

*18. Max-Planck-Institut für Astrophysik, Karl-Schwarzschild-Str. 1, D-85741 Garching bei München, Germany*

*19. INAF Osservatorio Astronomico di Trieste, Via Tiepolo 11, I-34131 Trieste, Italy*

*20. Computing & Minor Planet Section (Center for Astrodynamics) of the Oriental Astronomical Association, Sumoto, Japan*

*21. Department of Physics, Stockholm University, AlbaNova University Center, SE-10691 Stockholm, Sweden*

**The death of massive stars produces a variety of supernovae, which are linked to the structure of the exploding stars[1,2]. The detection of several precursor stars of Type II supernovae have been reported (e.g. Ref. 3), however we do not yet have direct information on the progenitors of the hydrogen deficient Type Ib and Ic supernovae. Here we report that the peculiar Type Ib supernova SN2006jc is spatially coincident with a bright optical transient[4] that occurred in 2004. Spectroscopic and photometric monitoring of the supernova leads us to suggest that the progenitor was a carbon-oxygen Wolf-Rayet star embedded within a helium-rich circumstellar medium. There are different possible explanations for this pre-explosion transient. It appears similar to the giant outbursts of Luminous Blue Variables (LBV) of 60-100 solar mass ($M_\odot$) stars[5], however the progenitor of SN2006jc was helium and hydrogen deficient. An LBV-like outburst of a Wolf-Rayet star could be invoked, but this would be the first observational evidence of such a phenomenon. Alternatively a massive binary system composed of an LBV which erupted in 2004, and a Wolf-Rayet star exploding as SN2006jc, could explain the observations.**

SN2006jc was discovered in UGC4904 on 2006 Oct 9.75 UT at magnitude 13.8 (Ref 4). The early spectrum was that of a hydrogen-poor event with strong, narrow HeI emission lines[6-9] superimposed on a broad-line spectrum of a Type Ic supernova. In 2004, an optical transient was reported to the Central Bureau for Astronomical Telegrams (CBAT) which, when retrospectively compared with the SN2006jc discovery images, appeared to be spatially coincident with the new supernova[4]. The 2004 transient was much fainter than SN2006jc (magnitude $\simeq$ 18) and remained visible only for few days after discovery. This event was never independently confirmed, and CBAT did not issue an official object designation. Given the new, bright SN discovery, the nature of this transient (which we name UGC4904-V1, i.e. variable 1 in UGC4904) has become intriguing. We aligned the images containing the two transients using differential astrometry of 21 nearby stars and find that UGC4904-V1 and SN2006jc are indeed coincident to within the uncertainties (Figure 1). Supplementary Information contains a comprehensive description of the method, the error analysis and a demonstration that the probability that these two events are chance coincidences is negligible.

We monitored the light curve and spectral evolution of SN2006jc during the first ~2 months after discovery (Figures 2 and 3). An independent dataset was presented by Ref.10. The object was discovered a few days past maximum, but still its brightness is comparable with that of the most luminous Ic supernovae (Figure 2), namely $M_R <$ -18.3, adopting a host galaxy distance of 25.8 $\pm$ 2.6 Mpc and a total reddening E(B-V) = 0.05 (see Supplementary Information). However, SN2006jc declines faster than most Type Ib/c supernovae and the optical spectra are unusual. The broad emission lines commonly observed in Type Ic supernovae are detected, although with an atypical profile (see Supplementary Information). In addition, prominent and relatively narrow (FWHM ~ 2,200 km s$^{-1}$) HeI emission lines are observed, hence the classification of SN 2006jc as a peculiar Type Ib event[6-10]. Although very rare, objects like SN 2006jc have been observed before: SN1999cq[11] and SN2002ao[10,12] (Figure 2b). The detection of narrow HeI lines suggests that SN2006jc-like events should be more properly classified as Type Ibn, in analogy to the similar nomenclature (Type IIn) used for supernovae which show narrow hydrogen emission lines. Another property of the spectra of SN2006jc (in common with SN1999cq and SN2002ao) is the blue colour, which remains almost constant over the entire observational period. The blue spectral continuum, the presence of narrower lines superimposed on broad spectral features (Figure 3) and the strong X-ray emission[13] are normally interpreted as a signature of interaction between supernova ejecta and circumstellar medium (CSM). The presence of prominent HeI lines and the simultaneous lack of conspicuous hydrogen features suggest a helium-rich composition of the CSM, while the abundance of hydrogen must be modest. There is no evidence of broad helium components, suggesting that the progenitor of SN2006jc had entirely lost its helium envelope and was a carbon-oxygen Wolf-Rayet star (WC or WO)[10]. Fast-moving ejecta produce the broad lines of intermediate-mass elements, whose width increases with time (from about 4,000 to 9,000 km s$^{-1}$), while the slow-moving CSM produces moderately narrow (about 2,200 km s$^{-1}$) HeI emissions. In addition to these prominent features, weak, very narrow (~500 km s$^{-1}$) P-Cygni absorptions attributed to HeI and OI (and, possibly, H$\alpha$) are visible in the highest resolution spectra, and are indicative of further undisturbed, slowly-moving shells originating from previous mass loss episodes (Supplementary Information). Severe mass loss is necessary to remove the outer helium and hydrogen layers, and to produce a massive carbon-oxygen core.

The 2004 outburst of UGC4904-V1 reached a peak magnitude of $M_R$ ~-14.1, and we offer a few possible explanations of the event. Giant outbursts of Luminous Blue Variables (LBVs)[14,15] are well documented transients with a similar peak magnitude and sharp decline (Figure 4 and Supplementary Information). LBVs are massive blue stars that show significant optical variability, due to unstable atmospheres and episodic mass loss. The only Galactic star well observed during such an eruption is $\eta$Carinae (in 1837-1857[14]), which has a current mass of approximately 90M$_\odot$ and initial mass of around 150M$_\odot$[16]. Other LBVs in the Local Universe that have shown giant outbursts of similar magnitude to UGC4940-V1 are likely to have had initial masses in the range 60-100M$_\odot$ (Supplementary Information). However, despite the magnitude of the 2004 outburst of UGC4904-V1 was similar to that of a typical LBV, an LBV scenario raises two problems. It is at odds with current stellar evolutionary theory which predicts that massive stars do not undergo core-collapse in the LBV stage, and the subsequent Wolf-Rayet star should have a lifetime of more than 200,000 years[1,2]. Additionally all the known LBVs which have undergone outbursts still have hydrogen and helium rich atmospheres[14,16] (Supplementary Information). The progenitor of SN2006jc was different, since prominent H lines were not detected in early supernova spectra (Figure 3). One could then propose that the progenitor star has been a Wolf-Rayet star for this time-scale, and the 2004 event was an LBV-like eruption of a Wolf-Rayet star[10]. In this case we need to invoke a novel explosion mechanism as no carbon-oxygen star has ever been observed to produce such a bright outburst.

It is unquestionable that SN2006jc is not a typical supernova, and hence this post-LBV channel is probably not the preferred one which produces Ib or Ic supernovae and the rarity of such events could be due to the high progenitor mass. If this scenario is true, then it has interesting implications. A star of 60-100$M_\odot$ has a carbon-oxygen core of mass 15-25$M_\odot$ as it enters the Wolf-Rayet phase[1,2] (Supplementary Information). The core-collapse of such an object is predicted to form a black hole by fall-back, producing a low yield of $^{56}$Ni and hence resulting in a very faint and under-energetic explosion[17]. However, although powered by the ejecta-CSM interaction, SN2006jc is a high-luminosity event and a plausible model for the production of a bright supernova from a black hole forming core is a jet-powered supernova[2,17,18]. Such events in He and H-free stars are the working models for long duration gamma-ray bursts[17].

As an alternative to the single star scenario, one could propose a massive binary system with two stars entering the final, violent, stages of their evolution. One of the components could have undergone a classical LBV outburst in 2004, while the companion was an evolved Wolf-Rayet star that collapsed to give SN2006jc. The interaction of the ejecta within a complex, circumstellar environment shaped by the strong stellar winds of massive stars could explain the numerous gas shells detected in the spectra (Supplementary Information). It has recently been suggested that $\eta$Carinae has a hot companion (around 30M$_\odot$) which is a nitrogen rich late O-type or Wolf-Rayet star[19]. A similar scenario (a pair of 30M$_\odot$ and 50M$_\odot$ stars) was proposed for HD5980, an LBV + WR star binary in the Small Magellanic Cloud[20]. In neither of these cases is the WR star a WC or WO, as it is sill nitrogen and helium rich, however a more evolved system is a possible progenitor. Also the system has only undergone moderate amplitude outbursts, reaching $M_V$=-10.6, significantly fainter than UGC4904-V1. Theoretical models of pre-supernova evolution in massive binary systems show that Type Ib/c supernovae can be produced, and a higher mass system than that calculated in Ref. 21 could be plausible, although a physically consistent scenario needs a detailed calculation.

Further observational and theoretical study are required to determine which scenario is the more likely. The detection in very late spectra of SN2006jc of more prominent, narrow H lines from the CSM (Mattila et al. in preparation) might support a single, massive star scenario. Moreover, line profile measurements may determine the asphericity of the ejecta and explosion[22], and radio modulations could eventually show that an LBV phase recently occured[23]. Finally, deep high resolution images with the Hubble Space Telescope can probe the presence of the possible binary companion, and a detailed comparison with few similar supernovae[10,11] may help to provide additional clues for the progenitor scenario.

**Supplementary Information** accompanies the paper on **www.nature.com/nature**.

Acknowledgements and the Competing Interests statement:

This work, conducted as part of the award "Understanding the lives of massive stars from birth to supernovae" made under the European Heads of Research Councils and European Science Foundation EURYI Awards scheme, was supported by funds from the Participating Organisations of EURYI and the EC Sixth Framework Programme, and also the Leverhulme Trust. JD is supported by the NSFC Grant No. 10673014. This paper is based on observations collected at the Asiago Observatory (Italy), the 2.16m Telescope of National Astronomical Observatories (China), the Observatoire de Haute-Provence (France), Telescopio Nazionale Galileo, Nordic Optical Telescope, Liverpool Telescope and William Herschel Telescope (La Palma, Canary Islands, Spain). We thank the support astronomers working at the Liverpool Telescope and Telescopio Nazionale Galileo for performing the follow-up observations of SN 2006jc. We are grateful to M. Ganeshalingam, A. V. Filippenko, R. J. Foley and W. Li for providing us the photometric data of SN 2002ao.

Correspondence and requests for materials should be addressed to Andrea Pastorello (a.pastorello@qub.ac.uk).

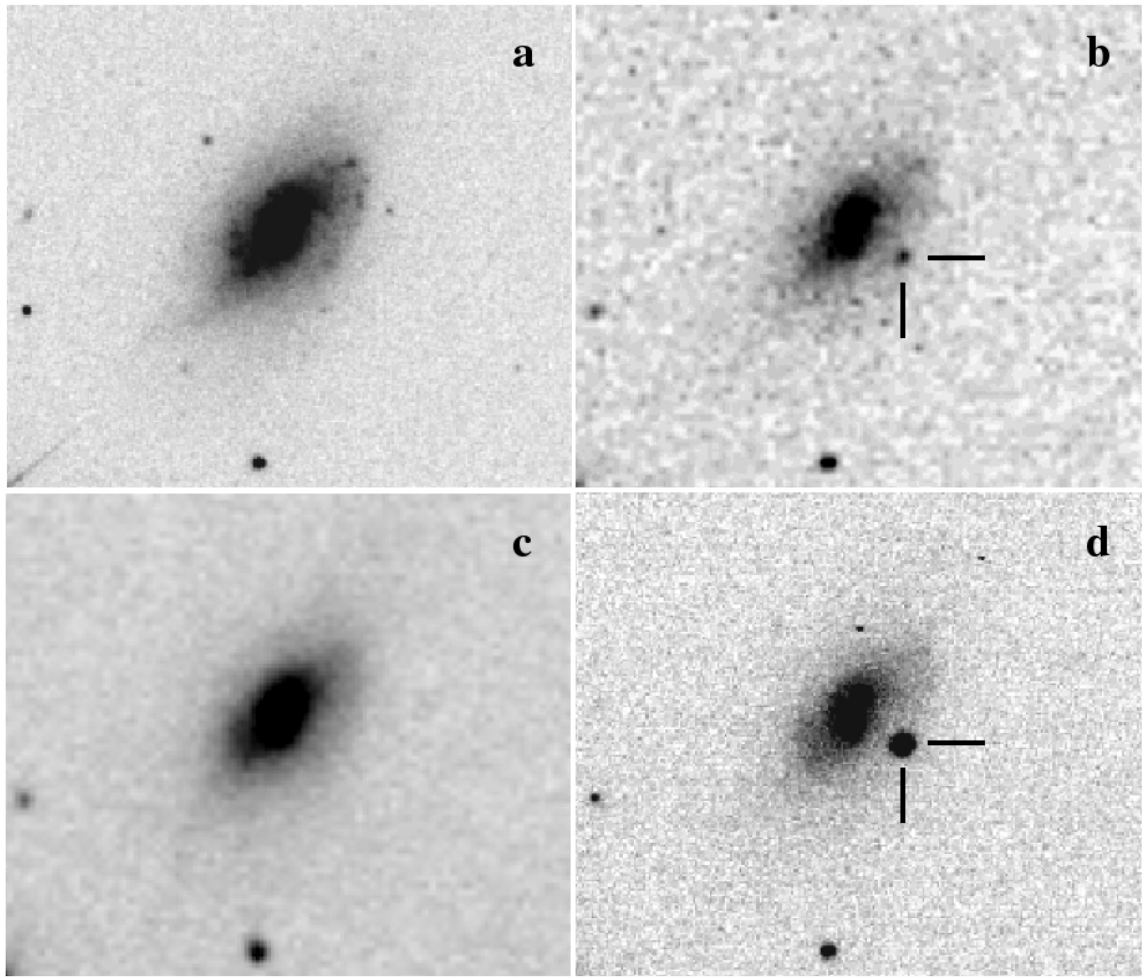

**Figure 1 | The transients UGC4904-V1 and SN2006jc.** Sequence of images of UGC4904 rebinned to a pixel scale of 0.53 arcsec. (**a**) *r'* band image of UGC4904 from the Sloan Digital Sky Survey obtained on 2001, Dec 20. No transient is visible. (**b**) Detection of UGC4904-V1 on 2004, Oct 16 by Itagaki (magnitude = 19.13 ± 0.19), using a 0.60-m f/5.7 reflector and Bitran-CCD (Kodak KAF 1001E). The transient was detected in 5 epochs between 2004, Oct 14 and 2004, Oct 23. The original image has a pixel size of 1.44 arcsec, and seeing of 2.2 arcsec. (**c**) Another image from Itagaki (2006, Sep 21) showing no transient detection. (**d**) R band frame (original pixel scale of 0.473 arcsec, seeing of 2.0 arcsec) taken on 2006, Oct 29 with the Asiago 1.82m Telescope equipped with AFOSC. We find that the two transients are

coincident to within 0.1 arcsec, and the total error budget (including the uncertainty in the position measurements and the error of the geometric transformation) is 0.3 arcsec (see Supplementary Information). The transient UGC4904-V1 was not detected before 2004 Oct 01 or after 2004 Nov 07, and it is not a moving object, as there is no apparent motion between the 5 epochs in which it was detected.

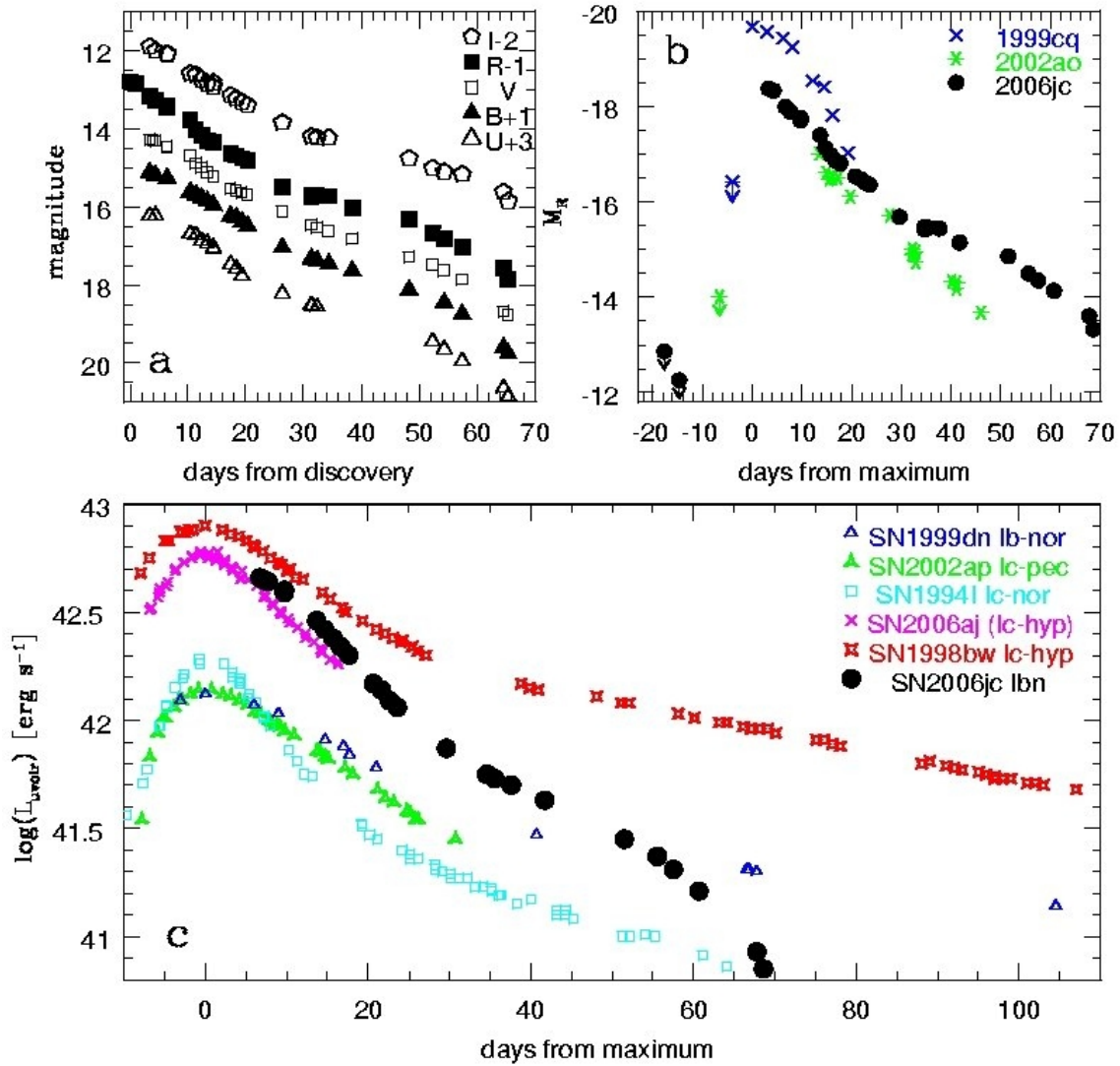

**Figure 2 | Light curve of SN2006jc.** (**a**) *UBVRI* light curves of SN2006jc. No significant colour evolution is visible from the available multiband photometry. (**b**) The R-band absolute light curve of SN2006jc is compared with unfiltered light curves of two similar interacting Type Ibn events: SN1999cq and SN2002ao (Refs. 10, 11, 24). The phases are estimated from the epochs of the approximated light curve maxima. The pre-explosion limit of SN1999cq was very close (~4 days) to the discovery epoch, which strongly constrains the epoch of this explosion and suggests (at least for this supernova) a very steep

rise to maximum light, supporting the idea that the ejecta are strongly interacting with a CSM. (**c**) Comparison between the quasi-bolometric light curve of SN2006jc and those of a sample of H-deficient core-collapse supernovae: 1999dn (normal SNIb, Benetti et al. in preparation), 1994I (normal SNIc, Ref. 25),  2002ap (high-velocity, moderate luminosity SNIc, Ref. 26), and the hypernovae 2006aj and 1998bw (either associated with an X-ray flash or a GRB; Refs. 27, 28, 29 and references therein). The light curves of SN2006jc and hypernova SN2006aj peak in between the luminous hypernova SN1998bw and more normal Type Ib/c supernovae, although the ejecta-CSM interaction might power significantly the observed light curve of SN2006jc.

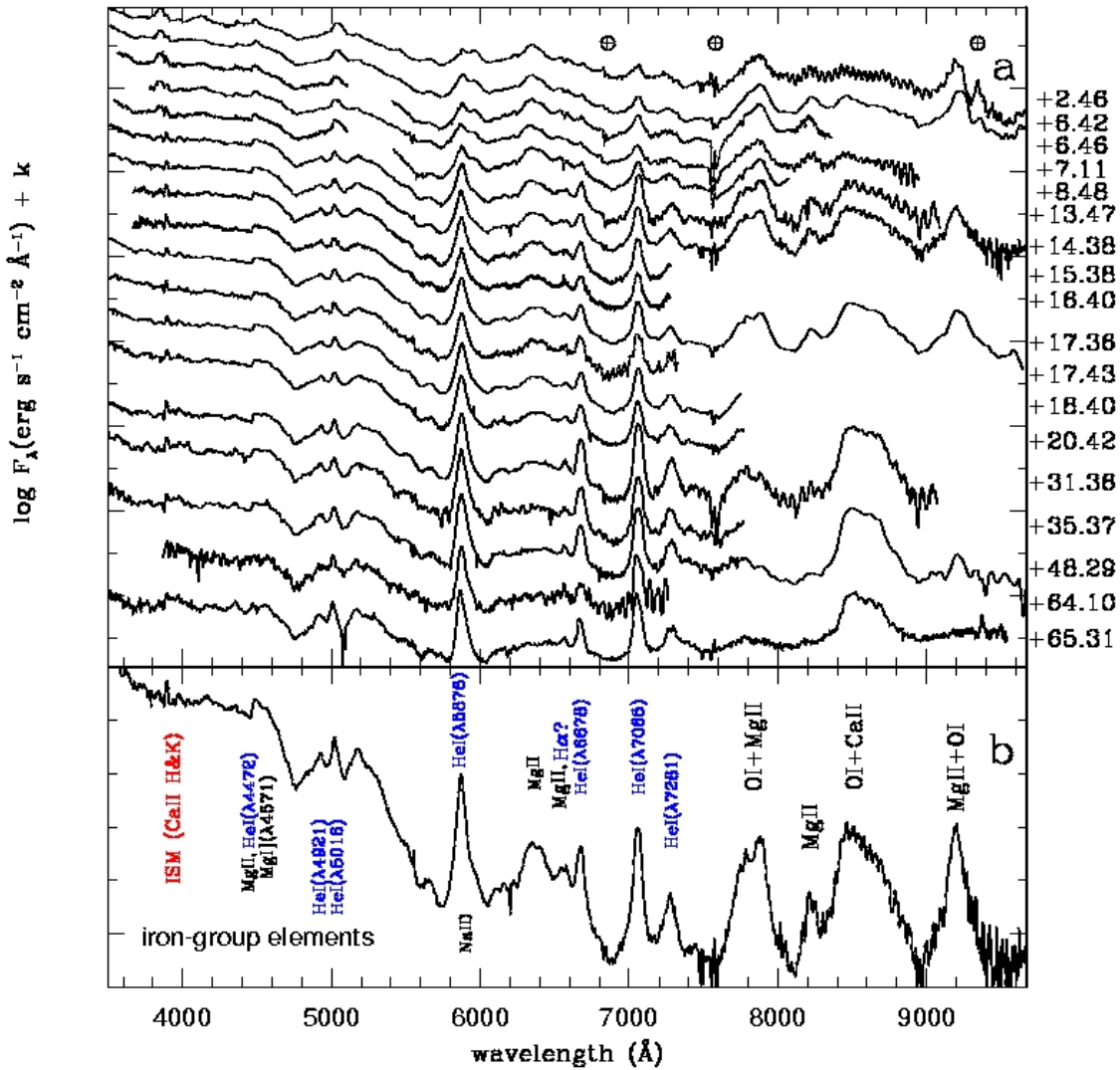

**Figure 3 | Spectra of SN2006jc and line identification.** Spectral evolution of SN2006jc (**a**) and identification of the main features in the spectrum at +14.38 days (**b**). All spectra have been shifted to the host galaxy rest wavelength. The spectra are dominated by a blue pseudo-continuum and broad lines (FWHM about 4,000 to 9,000 km s$^{-1}$, depending on the phase; black labels) of intermediate mass elements: NaID, OI $\lambda$7,774 and OI $\lambda$9,264 (both blended with MgII), CaII IR triplet (blended with OI $\lambda$8,446) and a number of MgII lines (the most prominent being $\lambda$4,481; $\lambda$6,347; $\lambda$6,546; $\lambda\lambda$7,877-7,896; $\lambda$

$\lambda$8,214-8,235; $\lambda\lambda$9,218-9,244). In addition the spectra show narrower (FWHM~2,200 km s$^{-1}$) HeI emissions (blue labels), strengthening with time. Narrow, weak interstellar CaII H&K lines are also visible (labeled in red). Moreover, the earliest spectra show some narrow unidentified emission lines in the blue region (e.g. at 3,850Å), and a weak, narrow H$\alpha$ absorption is possibly detected, replaced by a weak component in pure emission in the latest spectra. This detection suggests that a small amount of hydrogen is present in the He-dominated CS environment of SN2006jc. The nature of the flux excess in the blue region is unclear, possibly linked with the ejecta-CSM interaction (as observed in many Type IIn supernovae, e.g. SN1997cy), but might also result from a strong contribution from FeII lines (Ref. 10).

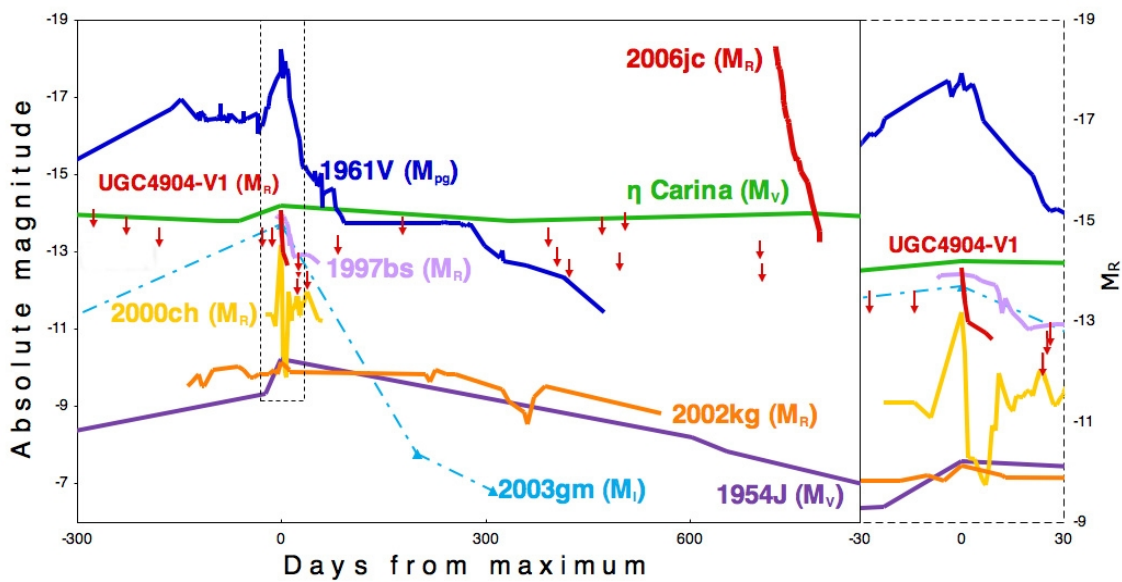

**Figure 4 | R-band light curves of UGC4904-V1 and SN2006jc compared with those of giant outbursts of LBVs.** The figure (without SN2006jc) was first shown in Ref. 15 (references therein) and supplemented with Ref. 30. Although all of the extragalactic transients were originally given supernova labels, they are now commonly accepted as giant outbursts of LBVs and not core-collapse supernovae. SN2002kg (which is the well known variable NGC2403-V37) was not a giant outburst, but is part of fairly common S Doradus variability of LBVs. The down-arrows are upper limits for UGC4904-V1. Due to many of the images of UGC4904 having limiting magnitudes in the range 19-20, we have only detected the peak of the outburst. With a peak magnitude of about $M_R = -14.1$ and such a sharp decline, a plausible explanation for UGC4904-V1 is a giant LBV-like outburst (Supplementary Information), although alternatives are presented in the text. The magnitudes of each object shown in figure are in different bands, depending on the available data.

# Supplementary Information for "A giant outburst two years before the core-collapse of a massive star" (Pastorello et al.)

1. Supplementary Methods

1.1 Astrometry

Accurate, relative astrometric calibration of the images obtained during the pre-explosion outburst and after the supernova explosion is critical for deciding on the positional coincidence of these two phenomena. The image obtained on 2006 Sep 21 by K. Itagaki (0.60-m f/5.7) was selected as the reference image as it is of good image quality and contains no trace of either UGC4904-V1 or SN2006jc. Four images containing UGC4904-V1 and two images containing SN2006jc were selected to be aligned to this reference image. Using the IRAF APPHOT package, we measured the centroid position for 21 bright stars visible in all seven images. From this, using the IRAF package GEOMAP, we derived geometric transformation functions for each of the images. In order to employ a general non-linear transformation, we selected a 2nd-order polynomial model with half cross terms to calculate the geometric alignment function. Finally the images were registered to the reference image using the IRAF package GEOTRAN.

The accurate measurement of the coordinates of UGC4904-V1 in the aligned images proved-out to be difficult due to the high background gradient from the underlying host galaxy. To remove the host galaxy contamination we subtracted the reference image from each of the aligned images prior to the position measurements. For this, the reference image was convolved to match the seeing conditions of the images containing UGC4904-V1, and the image intensity levels were matched using the Optimal Image Subtraction[31,32] method implemented in the ISIS2.2 package. For deriving the image matching parameters we selected 24x24 pixel regions around ten isolated stars nearby to UGC4904-V1. First order kernel variability was used to compensate for the effects of possible PSF variability and uncertainties in the image alignment, and the background was modeled using a 1st or 2nd order polynomial surface. The same image matching and subtraction procedures were also carried out for the two unfiltered images of 2006, Oct 8-9 containing SN2006jc.

The positions of UGC4904-V1 and SN2006jc were measured in the different aligned and subtracted images making use of three different methods, namely centroiding, Gaussian fitting, and optimal filtering, all implemented in the IRAF APPHOT package. The resulting pixel coordinates and the mean coordinates for each epoch are listed in Supplementary Table 1. The errors in the image alignment can be estimated from the RMS of the residuals from fitting the transformation function to the data points and are listed in the table. Average image alignment errors in *x/y* estimated in this way are 0.03/0.10 and 0.04/0.09 pixels for the UGC4904-V1 and the SN2006jc images, respectively. For UGC4904-V1 it was clear that the dominating source of uncertainty was the position measurement of the transient itself, whereas for the significantly brighter SN2006jc, the uncertainties from the position measurement and the image alignment were of a similar magnitude. We hence obtained the best estimates for the positions of UGC4904-V1 and SN 2006jc from a mean of the individual measurements, and adopted the standard error of the mean (SEM) as the uncertainty. For UGC4904-V1 we took the mean of the 12 individual measurements and the SEM is the standard deviation of these divided by $\sqrt{4}$ (the number of independent images). For SN 2006jc the standard deviation was calculated from the six individual measurements and the SEM calculated assuming two degrees of freedom, because we have only two independent images. These uncertainties include both the errors due to the image alignment and the actual position measurements for the outburst and the supernova. The final, average pixel coordinates of UGC4904-V1 were hence (472.68±0.12, 430.28±0.15) and for SN2006jc we found (472.66±0.06, 430.35±0.10). Thus the positions of UGC4904-V1 and SN2006jc are precisely coincident to within the uncertainties in the alignment technique which has a total uncertainty of 0.22 pixels. Given the pixel scale of 1.44 arcsec/pixel, the position of the transients formally differs by 0.1± 0.3 arcsec.

**1.2 Data Reduction**

All basic steps of data reduction were performed using tasks in the IRAF environment. Both photometric and spectroscopic images were pre-processed with the standard IRAF task CCDPROC (including bias, overscan, flat field corrections, and trimmed to the useful physical detector area).

Instrumental magnitudes of SN 2006jc have been measured with the PSF-fitting technique using a custom built DAOPHOT-based package

(SNOoPY). This method provides reasonable results in the case of SN2006jc because the object was largely dominant over the host galaxy background. However in order to properly estimate the magnitudes of the transient UGC4904-V1 (which was significantly fainter than SN2006jc), we needed to remove the host galaxy contribution from the transient image using the numerous templates collected over the previous 2 years, when no transient source was visible at the current SN location. The subsequent transformation of instrumental magnitudes into standard Johnson-Bessell magnitudes was performed observing a number of photometric standard fields[33] during the same nights as the supernova observations. The final photometric zero-points were estimated for all nights comparing the magnitudes of a sequence of stars in the field of UGC4904 to the average magnitudes of the same stars obtained in a few, selected photometric nights. The unfiltered magnitudes of the transient UGC4904-V1 can be best compared to R-band magnitudes (due to the response curve of the CCD used for those observations), and are hence zero-point calibrated using the R-band average magnitudes of the same sequence of stars used to calibrate the magnitudes of SN2006jc. The complete photometric measurements of the two transients, UGC4904-V1 and SN2006jc, are reported in Supplementary Table 2.

Spectra of SN2006jc were obtained from the pre-processed frames by performing optimal extraction with the IRAF task APALL, and then wavelength calibrated using spectra of comparison lamps taken with identical instrumental configuration. The spectra were then flux calibrated using sensitivity functions derived from spectra of standard stars obtained during the same night as the supernova observation. The spectra of the same stars have been used also to remove the most offending telluric absorption bands from the supernova spectra. Finally, the supernova spectra were checked with the photometry and, in cases of discrepancy, the spectra were adjusted to match the photometric data. The log of the spectroscopic observations presented in this paper is in Supplementary Table 3.

## 2. Supplementary Discussion

### 2.1 The probability of a chance spatial coincidence

One could conjecture that the two events UGC4904-V1 and SN2006jc are spatially coincident by chance and that they are physically unrelated objects. This appears to us a particularly unlikely alternative as they both are rather rare events. If UGC4904-V1 was actually in UGC4904 then it is a very rare occurrence, as such LBV giant outbursts are uncommon, and a supernova with such a faint visual magnitude combined with a sharp decline rate (1.6 mags per 10 days) as seen in Fig. 2 (main manuscript) has never before been recorded. Also SN2006jc is a very peculiar supernova, and its spectral characteristics are very unusual. We can present some simple statistical arguments that strongly disfavour the spatial coincidence of two physically unrelated transients.

SN2006jc is certainly in the host UGC4904, as we see the absorption features due to the interstellar CaII H & K lines, at the same redshift as the optical nebular emission lines of the galaxy. If UGC4904-V1 is also in the galaxy it must have been either a faint supernova or a giant outburst of an LBV. Classical novae only reach magnitudes of $M_R \approx -10$ (Ref. 34), and there is no other known transient that could reach that magnitude. Hence in any one galaxy, we should determine the probability that two supernovae, or more correctly a supernova and LBV outburst, are spatially coincident. If we assume the rate of supernovae and LBV outbursts in any particular galaxy are the same at 1 per 100 yrs, and that typical nearby supernova searches (such as KAIT and most well equipped amateur efforts) typically have been working for 10 years, then the Poissonian probability that 2 or more supernovae (or LBVs, or any combination thereof) occur within a period of 10 years is $p_{10}$=0.0047.

At the distance of UGC4904 (25.8Mpc), a 0.3″ radius around SN2006jc corresponds to an area of 4300 pc$^2$. If we assume that the probability of a coincidental and unrelated event *P(coincidence)* within this area of a galaxy with typical radius of 5 kpc is simply the ratio of the areas times the event rate, then

*P(coincidence)=(5.5×10$^{-5}$× p$_{10}$)/ sini*

where sin*i* corrects for the projection effects of galaxy inclination (*i = angle of inclination).* If we assume randomly inclined axes, and hence on average sin*i = 0.7,* then P(coincidence)=3.7x10$^{-7}$. In other words, in the 10 years of focused searching for nearby supernovae the probability that any one, typical, galaxy has had two supernovae (or a supernova and an LBV) *and* that they are spatially coincident is 3.7x10$^{-7}$. If we assume a sample of 10000 galaxies are monitored by such supernova surveys[35], it would suggest a total probability that 2 supernovae are detected in any galaxy and are spatially coincident to be *P(total)~*3.7x10$^{-3}$. Admittedly we have used some general assumptions. We assumed a mean inclination angle, a mean supernova rate for galaxies, and assumed that events would be randomly distributed in a galaxy. The latter assumption could be compromised, as massive star explosive transients will obviously be more concentrated in the high star formation areas. The SN rate of 1 per 100 yrs is, within a factor of two, what we tend to observe locally[36]. On the other hand we have conservatively assumed that all supernovae that occur are actually discovered, and relaxed the coincidence time to 10 years rather than 2. In fact the probability that two supernovae are actually discovered in the same galaxy is significantly less than the theoretical Poissonian probability that they occur. To address these assumptions would require a detailed Monte Carlo simulation, for which the uncertainties in the sampling frequencies and the depths of observations from all the supernova surveys would have to be included, and at present they are not known accurately. Also the rate of LBV giant outbursts appear to be significantly lower than the rate of supernovae, as only ~5% of recent supernova candidate transients are classified as LBV outbursts[37]. Although we have made assumptions that could be questioned, on balance we have been conservative and would expect *P(total)* to be lower than we have estimated.

If UGC4904-V1 is a foreground transient we can estimate probabilities of coincidence. We assume it was a faint classical nova (CN). It has a very rapid decay rate even for "fast-nova", which typically have decline rates of 0.5 mags per 10 days [34]. The "very fast" novae in M31 have decline rates similar to UGC4904-V1, so we assume conservatively that all novae in our galaxy are candidates. The rate of CNe in the Milky Way is $35\pm11$ yr$^{-1}$ (Ref. 38), and hence in 100 years there have been around $3,500\pm1,100$ CNe. In approximately 100 years of supernova searches there have been around 3,600 supernovae[35]. If we assume that CNe are isotropic, again a very conservative assumption as they will be concentrated in the disk and bulge (and UGC4904 has coordinates $l$=179° $b$= 44°), then the possibility of any CN being within our measurement uncertainty (0.3″ radius) of a single supernova is $6\times10^{-13}$, and hence around $2\times10^{-9}$ for any supernova in the sample of 3,600. The rate of dwarf novae in the galaxy is uncertain. However given the number known, it is unlikely to differ from the CN rate by more than an order of magnitude[39]. Any galactic transient which would have a probability of more than about 1% of being coincident with *any* of the 3,600 supernovae would have to be ~$10^7$ more common than CNe. We can think of none which are likely candidates. Clearly both of these illustrative calculations suggest a chance occurrence is very unlikely.

## 2.2 Properties of the host galaxy UGC4904: distance, reddening and metallicity

The distance to UGC4904 can currently only be estimated from the recessional velocity. The velocity of UGC4904 corrected for infall of the Local Group towards Virgo is 1,830 km s$^{-1}$ (from the HyperLeda catalogue[40]) and adopting a Hubble constant of $H_0$=71 km s$^{-1}$ Mpc$^{-1}$ gives a distance of 25.8 ± 2.6 Mpc, where the uncertainty comes from the cosmic thermal velocity dispersion[41] of 187 km s$^{-1}$.

We adopt a foreground Galactic extinction[42] toward UGC4904 of E(B-V)=0.02± 0.02. We do not have an estimate of the internal host galaxy reddening towards SN2006jc. However we believe that it is likely to be small for three reasons. The spectrum is very blue, rising steeply towards the UV and one can compare an arbitrary, and unphysical, hot blackbody temperature ($T_{BB}$) to the dereddened continuum[43]. If the earliest spectrum of SN2006jc is dereddened with a standard reddening law characterized with a value *E(B-V)*>0.5, then even a blackbody of $T_{BB}$=10$^9$ K cannot match the steepness of the bluewards rising continuum. However a $T_{BB}$ ~12,000K is more likely to be applicable, given the typical effective temperatures estimated for the expanding photospheres of core-collapse supernovae at early phases[44]. To approximately match the slope of a black body of $T_{BB}$ =12,000K, the spectrum of SN2006jc needs to be  dereddened by only *E(B-V)*=0.05. In addition, the detection of weak interstellar CaII H&K lines  in the supernova spectra, attributed both to UGC4904 and our Galaxy, with comparably narrow equivalent widths (~0.1Å), makes the total extinction necessarily small. Finally, the galaxy itself is faint and blue, and SN2006jc occurred well away from the central regions and hence it appears unlikely there is significant internal dust extinction. With these motivations, we have adopted in this paper a relatively small value for the total reddening, *E(B-V)*=0.05.

The host galaxy was observed spectroscopically in the Sloan Digital Sky Survey and the emission line strengths are available through DR5[45]. An estimate of the nebular oxygen abundance of UGC4904 can be determined from the [OIII] and [NII] line strengths (using calibrations discussed in Ref. 46) giving 12+log(O/H) = 8.30±0.11 dex (or a global metallicity of 0.5$Z_\odot$). The absolute magnitude of UGC4904 ($M_B$ ≈ -16.0) similar to the Small Magellanic Cloud, and hence a low metallicity is expected from the well established mass-metallicity relation[47]. Hence it appears that the environment of SN2006jc is mildly metal deficient, somewhat similar to the massive stars in the Large Magellanic Cloud, which has an oxygen abundance of 8.35±0.10 dex[48].

### 2.3 Might SN2006jc be an LBV-like super-outburst?

Some doubts on the real nature of SN2006jc may arise from the peculiar characteristics of its spectra. Together with the narrow HeI lines, the spectra have a very sharp, blue continuum and, at the red wavelengths, broad lines showing an unusual profile, which do not show the clear P-Cygni absorptions expected in a young Type Ib/c SN.

As an alternative scenario to a supernova explosion, one might claim that SN2006jc is a further, exceptional LBV-like outburst occurring 2 years after UGC4904-V1, and about 4 magnitudes more luminous than that event. The magnitude at maximum of SN2006jc is indeed similar to that observed in the brightest super-outburst observed so far, i.e. SN1961V, although the real nature of this event (super-outburst of an LBV or genuine supernova explosion) is still debated[49-51]. SN2006jc and SN1961V show a similar post-maximum decline of the light curve, but the rising phase to the maximum light is definitely slower in SN1961V, increasing by only ~2 magnitudes in about 40 days[52,53]. As shown in Supplementary Figure 1, if we adopt the light curve of SN1999cq[11] as a template, SN2006jc would have a very steep rise to the maximum, increasing by >3 magnitudes in about 4 days. Without invoking an analogy between the light curves of SN1999cq and SN2006jc, the deep detection limit obtained before the discovery of SN2006jc indicates that this event rose by about 6 magnitudes in about 20 days, so much more rapidly than SN1961V.

Moreover, there is no clear spectral similarity between SN2006jc and SN1961V. Spectra of SN1961V[53,54] are dominated by prominent, narrow H and HeI emission lines, and there is no evidence of any broad features. However, since the spectra of SN1961V lack a proper flux calibration, it is impossible to verify if they showed a strong, blue continuum, similar to that observed in SN2006jc. Therefore we find no definite similarity between SN2006jc and the putative super-outburst SN1961V.

On the contrary SN2006jc shows many features in common with Type Ic supernovae. Again, the maximum luminosity and the post-maximum decline of the light curve are not rather similar to those of the Type Ic SN2006aj (as shown in Figure 2), and the fast pre-maximum rise might be the result of strong interaction with a CSM, analogous to the fast rise to maximum occasionally observed in some Type IIn supernovae (e.g. SN1994W[55]). The similarity with a normal Type Ic becomes more evident if we compare the latest spectrum available for SN2006jc with that of the normal Type Ic SN1988L[56] (Supplementary Figure 2). Most of the broad spectral features are visible in both spectra, although the extremely blue continuum remains a peculiarity of SN2006jc and the few similar events[10,11], although already observed in several CSM-interacting supernovae (e.g. SN 1997cy[57] and SN 1999E[58]).

**Supplementary Figure 1 | Comparison of the absolute light curves of SN2006jc (and the similar SN1999cq and SN2002ao) with the putative LBV super-outburst SN1961V.**

Despite the peak luminosity and post-maximum light decline of SN2006jc are similar to those of SN1961V[52,53], its pre-maximum rise (as well as those of the two similar objects[10,11,24]) is much faster than in SN1961V, showing that these events are not alike.

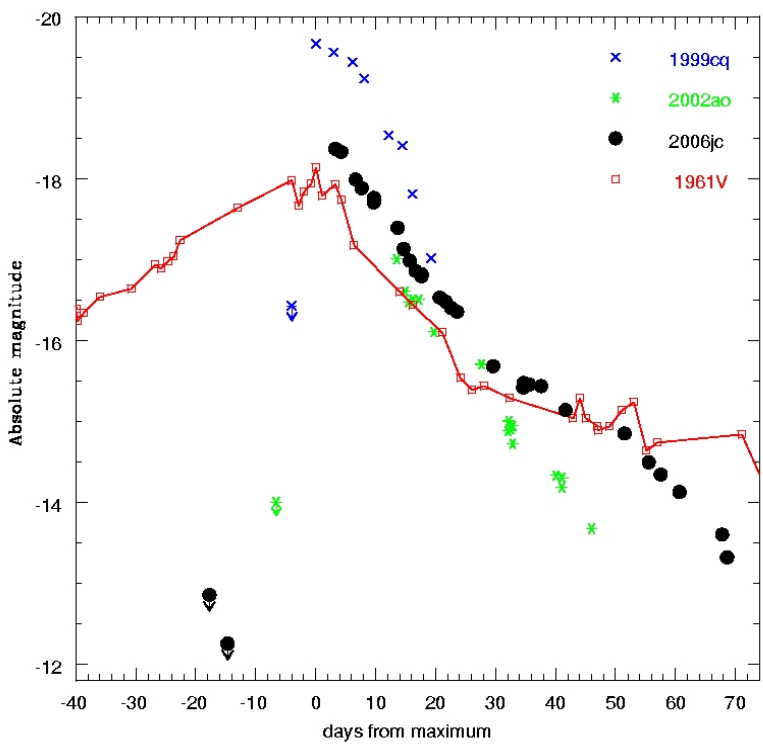

**Supplementary Figure 2: Comparison between a spectrum of SN2006jc and that of the normal, non-interacting supernova type Ic SN1988L approaching the nebular phase.**

The spectrum of SN2006jc is at phase +35.37 days from discovery; that of SN1988L (rescaled to the flux of the spectrum of SN2006jc) was obtained on June 06, 1988 and is shown in Reference 56. This spectrum shows the typical features of a Type Ic supernova during the transition between the photospheric and the nebular phase. The most important broad emission lines are marked with a long-dashed red line (and identified); the main absorption features are indicated with a short-dashed green line. Most of the broad features are visible in the two spectra. However, a remarkable difference is that the [OI] $\lambda\lambda 6,300$-$6,364$ and [CaII] $\lambda\lambda 7,291$-$7,324$ classical nebular emission featuress are still missing in the spectrum of SN2006jc, indicating at this epoch the supernova has not entered the nebular phase yet.

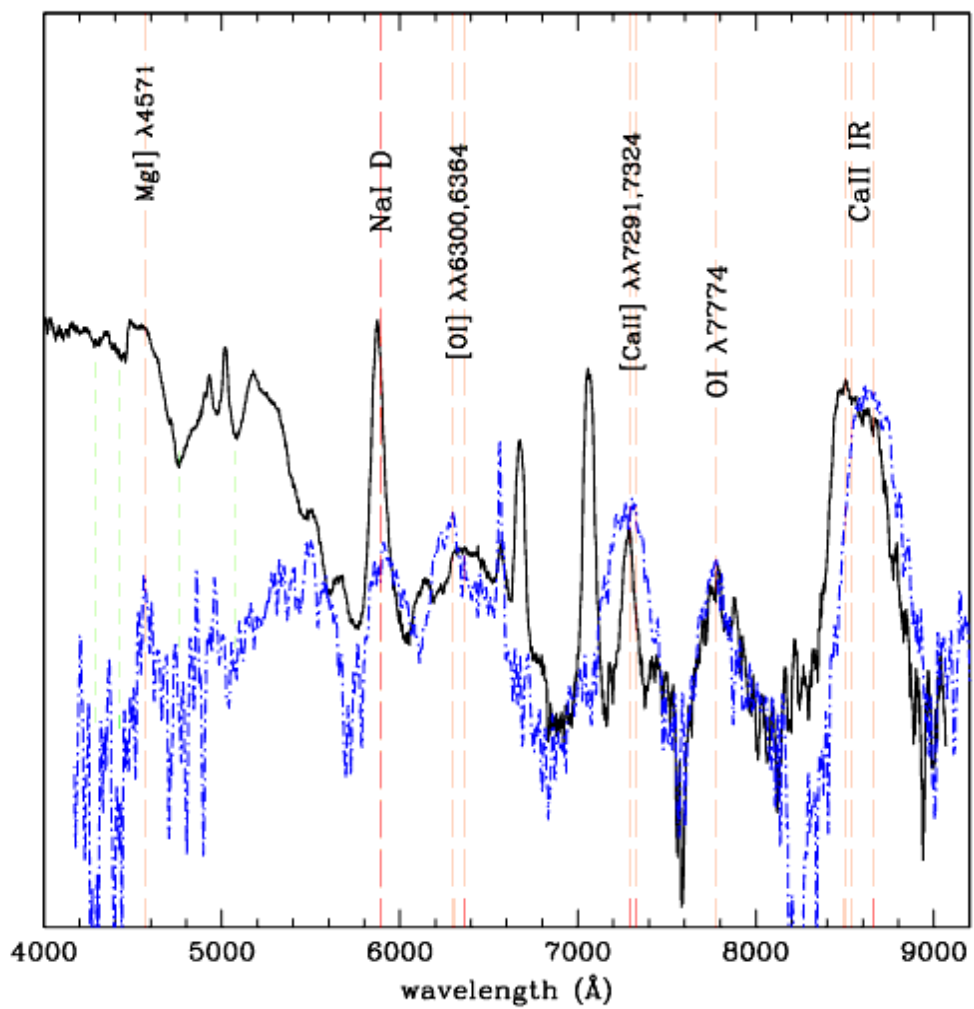

## 2.4 Evidence of circumstellar shells and possible relation with the 2004 outburst.

The prominent, moderately narrow (~2,200 km s$^{-1}$) HeI emission lines visible in the spectra of SN 2006jc give unequivocal evidence for the presence of a massive He-rich shell ejected by the progenitor of this supernova (see also Ref. 10). These lines are possibly produced by the interaction between the supernova ejecta (expanding at a velocity of 4,000-9,000 km s$^{-1}$) and a CSM expanding at a velocity of about 2,200 km s$^{-1}$. Such a shocked CS shell, moving at 2,200 km s$^{-1}$, could have been already reached by the supernova ejecta only if it was produced by a mass-loss episode closer in time than the 2004 outburst.

Alternatively, narrow circumstellar lines superimposed on the broad-lined supernova spectrum can also be produced in undisturbed CS gas photoionized by the initial UV/X-ray flash. If the impact between supernova ejecta and CSM is not occurred yet, it is still possible that this material has been produced in the 2004 outburst event. However, the unusual strength of the HeI lines and their long-life make very unlikely the photoionization mechanism.

Additional information comes from the two higher resolution WHT spectra, in which a second circumstellar component, visible as narrow, blue-shifted absorption lines of HeI and OI, was detected. Another narrow unresolved P-Cygni feature, probably H$\alpha$, is also detected in most of the early spectra. H$\alpha$ is also visible in net emission in the latest spectra. This suggests that some hydrogen is present in the CSM environment of SN2006jc. All these very narrow features are indicative of the presence of at least one further slow-moving (~500 km s$^{-1}$) circumstellar shell. Such narrow P-Cygni absorptions are not unusual, having been already detected in a number of Type IIn supernovae (e.g. 1994W[55], 1994aj[59], 1995G[60], 1996L[61], 1997ab[62], 1997eg[63]), and proving that undisturbed circumstellar shells are frequently observed in the environment of many core-collapse supernovae. Recurrent mass loss episodes forming the complex CSM of SN2006jc are also in a good agreement with a massive LBV (or post-LBV) scenario for the precursor of this supernova.

As an alternative to a multi-shell environment, we might invoke a clumpy CSM [64] having a moderate pre-interaction bulk velocity of ~500 km s$^{-1}$, in which the 2,200 km s$^{-1}$ HeI lines may be explained by the shock velocity inside individual

He-rich clumps. However, a detailed study of the structure of the circumstellar environment of SN2006jc is beyond the aim of this paper and it will be more extensively discussed in a forthcoming paper.

## 2.5 Stellar evolutionary models of 60-100$M_\odot$ stars

We have calculated stellar models of 60$M_\odot$ and 100$M_\odot$ stars using the STARS code[1,65], in particular to determine the carbon-oxygen core mass before core-collapse. As an example, the evolutionary track for a 100$M_\odot$ star is shown in Supplementary Figure 3. The mass-loss prescription used is that of scheme D in Ref. 1, which is not sporadic and episodic as we suggest occurred for the progenitor of SN2006jc. The mass-loss mechanism, which removes the outer hydrogen and helium envelopes could be episodic or steady, or a combination of both, but whatever it is will not significantly change the mass of the carbon-oxygen core left. As SN2006jc has the characteristic features of a Type Ic supernova, it is likely the progenitor was the bare carbon-oxygen core of this 60-100$M_\odot$ star. We estimate the mass of the core to be 15-25$M_\odot$ at the end of core carbon burning. With such a large core mass, the formation of a black hole is predicted rather than a neutron star[1,2,17]. This is in contrast to recent estimates of the ejected mass from the Type Ic SN2006aj. The light curves of these 2 events are reasonably similar, as shown in Figure 2, but Ref. 28 suggests only 2$M_\odot$ of ejecta for SN2006aj, and that this arose in the core of a star of original mass 20-25$M_\odot$. The largest amount of ejecta estimated for any Type Ic supernova so far is about 13$M_\odot$ for SN2003lw[66], which was estimated to originate in a star of original mass of around 40-50$M_\odot$. Similarly massive ejecta (about 11$M_\odot$) were obtained for SN1998bw, with a main-sequence mass of 40$M_\odot$[67]. The giant outburst of UGC4904-V1, similar to those observed for a number of LBVs (Figure 4; see also Supplementary Table 4). If the 2004 outburst and SN2006jc are events produced by the same progenitor star (and not e.g. by two different stars belonging to a binary system, see discussion in the main paper), the natural consequence is that the progenitor star must have been very massive, as there is no precedent for stars even in the 30-60$M_\odot$ range to show such luminosities in the LBV phase. Hence this is perhaps one of the largest cores we have witnessed undergoing core-collapse.

**Supplementary Figure 3 | Hertzprung-Russel Diagram (left) and interior structure (right) of a 100M$_\odot$ star.**

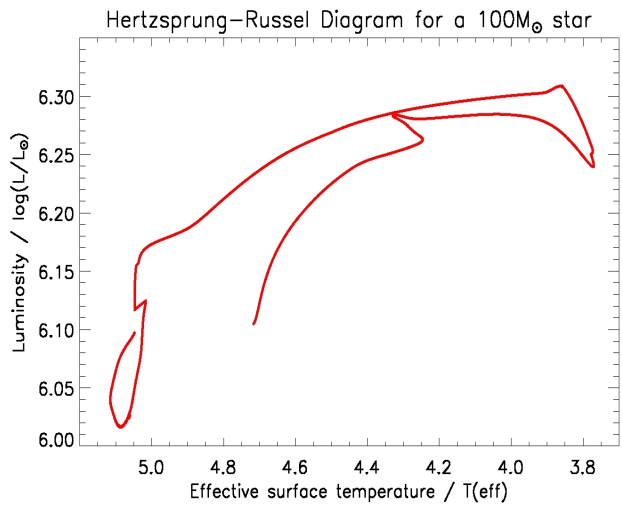
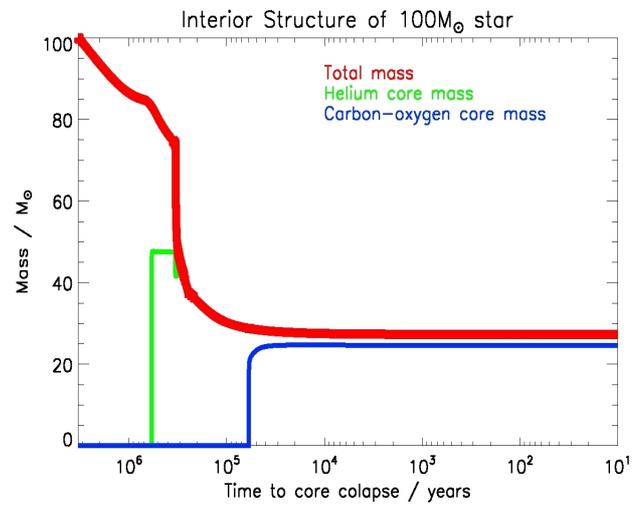

## 3. Supplementary Tables

**Supplementary Table 1: Position measurements for the outburst and the supernova.**

The positions of the transients from measurements using three different methods are listed. The *x* and *y* RMS of the residuals from fitting a transformation function between the data points of each epoch and the reference image are listed in the 6th and 7th column (in pixels).

| Epoch | Object | Centroid | Gauss | Ofilter | Mean | Fit $x_{RMS}$ | Fit $y_{RMS}$ |
|---|---|---|---|---|---|---|---|
| 2004 Oct 15 | UGC4904-V1 | 472.90, 429.84 | 472.87, 429.92 | 472.89, 429.92 | 472.89, 429.89 | 0.033 | 0.111 |
| 2004 Oct 16 | UGC4904-V1 | 472.57, 430.20 | 472.50, 430.28 | 472.55, 430.28 | 472.54, 430.25 | 0.025 | 0.102 |
| 2004 Oct 18 | UGC4904-V1 | 472.87, 430.51 | 472.85, 430.62 | 472.86, 430.62 | 472.86, 430.58 | 0.028 | 0.102 |
| 2004 Oct 22 | UGC4904-V1 | 472.40, 430.35 | 472.40, 430.41 | 472.42, 430.45 | 472.41, 430.40 | 0.023 | 0.103 |
| 2006 Oct 09 | SN2006jc | 472.69, 430.49 | 472.56, 430.43 | 472.54, 430.52 | 472.60, 430.48 | 0.043 | 0.085 |
| 2006 Oct 11 | SN2006jc | 472.74, 430.22 | 472.71, 430.22 | 472.71, 430.23 | 472.72, 430.22 | 0.044 | 0.086 |

**Supplementary Table 2 : Photometry of the outburst UGC4904-V1 and SN2006jc.**

Images from the 0.6m Itagaki telescope and from the 0.35m Boles reflector telescope were unfiltered and the response curve of the CCD used suggests that the unfiltered magnitudes are best compared to the R band photometry. Hence they are listed under the R column, although the difference should be recognised. The instruments used in the photometric observations of SN2006jc are the 3.58m Telescopio Nazionale Galileo, the 2.0m Liverpool Telescope and the 2.56m Nordic Optical Telescope in La Palma (Canary Islands, Spain) and the 1.82m Copernico Telescope of Mt. Ekar (Asiago, Italy).

| Date | JD | Telescope | U | B | V | R | I |
|---|---|---|---|---|---|---|---|
| 2004 Jan 12 | 2,453,016.52 | Boles 0.35m | | | | *>18.5* | |
| 2004 Mar 01 | 2,453,066.39 | Boles 0.35m | | | | *>18.7* | |
| 2004 Apr 17 | 2,453,113.44 | Boles 0.35m | | | | *>19.0* | |
| 2004 Sep 17 | 2,453,266.27 | Itagaki 0.6m | | | | >19.0 | |
| 2004 Sep 30 | 2,453,279.27 | Itagaki 0.6m | | | | >19.0 | |
| 2004 Oct 14 | 2,453,293.32 | Itagaki 0.6m | | | | 18.05 (0.23) | |
| 2004 Oct 15 | 2,453,294.31 | Itagaki 0.6m | | | | 18.73 (0.19) | |
| 2004 Oct 16 | 2,453,295.32 | Itagaki 0.6m | | | | 19.13 (0.19) | |
| 2004 Oct 21 | 2,453,300.33 | Itagaki 0.6m | | | | 19.35 (0.15) | |
| 2004 Oct 23 | 2,453,302.27 | Itagaki 0.6m | | | | 19.47 (0.17) | |
| 2004 Nov 07 | 2,453,317.22 | Itagaki 0.6m | | | | >20.2 | |
| 2004 Nov 08 | 2,453,318.34 | Itagaki 0.6m | | | | >19.8 | |
| 2004 Nov 09 | 2,453,319.26 | Itagaki 0.6m | | | | >19.6 | |
| 2004 Nov 22 | 2,453,332.17 | Itagaki 0.6m | | | | >20.1 | |
| 2005 Jan 05 | 2,453,375.58 | Boles 0.35m | | | | >19.2 | |
| 2005 Apr 10 | 2,453,471.46 | Boles 0.35m | | | | >18.7 | |
| 2005 Nov 10 | 2,453,684.60 | Boles 0.35m | | | | >19.0 | |
| 2005 Nov 24 | 2,453,699.31 | Itagaki 0.6m | | | | >19.5 | |

**Supplementary Table 2 : Photometry of the outburst UGC4904-V1 and SN2006jc.**

Continued.

| Date | JD | Telescope | U | B | V | R | I |
|---|---|---|---|---|---|---|---|
| 2005 Dec 11 | 2,453,716 | CBET 666 | | | | >19.8 | |
| 2006 Jan 29 | 2,453,764.67 | Boles 0.35m | | | | >18.7 | |
| 2006 Feb 24 | 2,453,791.12 | Itagaki 0.6m | | | | >19.7 | |
| 2006 Mar 03 | 2,453,798.38 | Boles 0.35m | | | | > 18.6 | |
| 2006 Sep 18 | 2,453,997.28 | Itagaki 0.6m | | | | > 19.3 | |
| 2006 Sep 21 | 2,454,000.30 | Itagaki 0.6m | | | | > 19.9 | |
| 2006 Oct 09 | 2,454,018.27 | Itagaki 0.6m | | | | 13.80 (0.03) | |
| 2006 Oct 10 | 2,454,019.28 | Itagaki 0.6m | | | | 13.84 (0.03) | |
| 2006 Oct 13 | 2,454,021.66 | LT + RATCAM | 13.24 (0.01) | 14.13 (0.01) | 14.28 (0.01) | 14.18 (0.01) | 13.91 (0.01) |
| 2006 Oct 14 | 2,454,022.68 | LT + RATCAM | 13.22 (0.01) | 14.19 (0.01) | 14.33 (0.01) | 14.29 (0.01) | 14.02 (0.01) |
| 2006 Oct 16 | 2,454,024.68 | NOT + ALFOSC | | 14.27 (0.01) | 14.48 (0.01) | 14.46 (0.01) | 14.11 (0.01) |
| 2006 Oct 16 | 2,454,024.70 | Ek1.82 + AFOSC | | 14.27 (0.02) | 14.46 (0.02) | 14.41 (0.03) | 14.08 (0.06) |
| 2006 Oct 20 | 2,454,028.65 | LT + RATCAM | 13.70 (0.01) | 14.62 (0.01) | 14.70 (0.01) | 14.78 (0.01) | 14.60 (0.01) |
| 2006 Oct 21 | 2,454,029.64 | LT + RATCAM | 13.73 (0.01) | 14.69 (0.01) | 14.90 (0.01) | 15.04 (0.01) | 14.63 (0.01) |
| 2006 Oct 22 | 2,454,030.66 | LT + RATCAM | 13.88 (0.01) | 14.78 (0.01) | 15.00 (0.01) | 15.19 (0.01) | 14.75 (0.01) |
| 2006 Oct 23 | 2,454,031.63 | LT + RATCAM | 13.94 (0.01) | 14.86 (0.01) | 15.13 (0.01) | 15.31 (0.01) | 14.86 (0.01) |
| 2006 Oct 24 | 2,454,032.63 | TNG + DOLORES | 14.08 (0.02) | 14.98 (0.01) | 15.23 (0.02) | 15.35 (0.01) | 14.83 (0.01) |
| 2006 Oct 24 | 2,454,032.63 | LT + RATCAM | 14.04 (0.01) | 14.96 (0.01) | 15.23 (0.01) | 15.38 (0.01) | 14.93 (0.01) |
| 2006 Oct 27 | 2,454,035.66 | Ek1.82 + AFOSC | 14.47 (0.01) | 15.23 (0.01) | 15.53 (0.01) | 15.64 (0.01) | 15.17 (0.02) |
| 2006 Oct 28 | 2,454,036.62 | Ek1.82 + AFOSC | 14.58 (0.01) | 15.27 (0.01) | 15.59 (0.02) | 15.69 (0.02) | 15.24 (0.01) |
| 2006 Oct 29 | 2,454,037.61 | Ek1.82 + AFOSC | | 15.40 (0.01) | 15.64 (0.03) | | |
| 2006 Oct 29 | 2,454,037.62 | Ek1.82 + AFOSC | 14.77 (0.01) | 15.39 (0.01) | 15.65 (0.01) | 15.77 (0.01) | 15.32 (0.01) |
| 2006 Oct 30 | 2,454,038.59 | Ek1.82 + AFOSC | | 15.49 (0.02) | 15.71 (0.03) | 15.82 (0.03) | 15.41 (0.02) |

**Supplementary Table 2 : Photometry of the outburst UGC4904-V1 and SN2006jc.**

Continued.

| Date | JD | Telescope | U | B | V | R | I |
|---|---|---|---|---|---|---|---|
| 2006 Nov 05 | 2,454,044.60 | LT + RATCAM | 15.22 (0.07) | 16.05 (0.08) | 16.13 (0.03) | 16.49 (0.03) | 15.86 (0.02) |
| 2006 Nov 10 | 2,454,049.58 | NOT + ALFOSC | 15.53 (0.03) | 16.33 (0.02) | 16.47 (0.02) | 16.75 (0.02) | 16.21 (0.01) |
| 2006 Nov 10 | 2,454,049.64 | LT + RATCAM | 15.55 (0.01) | 16.36 (0.01) | 16.47 (0.01) | 16.70 (0.01) | 16.21 (0.01) |
| 2006 Nov 11 | 2,454,050.58 | LT + RATCAM | 15.57 (0.05) | 16.37 (0.03) | 16.54 (0.06) | 16.72 (0.07) | 16.24 (0.06) |
| 2006 Nov 13 | 2,454,052.59 | Ek1.82 + AFOSC | | 16.46 (0.06) | 16.63 (0.06) | 16.74 (0.08) | 16.25 (0.07) |
| 2006 Nov 17 | 2,454,056.67 | LT + RATCAM | | 16.64 (0.01) | 16.83 (0.01) | 17.03 (0.01) | |
| 2006 Nov 26 | 2,454,066.47 | Ek1.82 + AFOSC | | 17.13 (0.02) | 17.29 (0.02) | 17.32 (0.03) | 16.77 (0.02) |
| 2006 Dec 01 | 2,454,070.53 | LT + RATCAM | 16.46 (0.04) | | 17.49 (0.02) | 17.68 (0.03) | 17.02 (0.03) |
| 2006 Dec 03 | 2,454,072.52 | LT + RATCAM | 16.66 (0.07) | 17.46 (0.16) | 17.64 (0.12) | 17.83 (0.06) | 17.11 (0.06) |
| 2006 Dec 06 | 2,454,075.65 | LT + RATCAM | 16.96 (0.05) | 17.74 (0.03) | 17.87 (0.03) | 18.04 (0.05) | 17.18 (0.03) |
| 2006 Dec 13 | 2,454,082.77 | TNG + DOLORES | 17.64 (0.02) | 18.61 (0.02) | 18.69 (0.03) | 18.57 (0.03) | 17.63 (0.04) |
| 2006 Dec 14 | 2,454,083.58 | LT + RATCAM | 17.86 (0.10) | 18.77 (0.05) | 18.78 (0.05) | 18.85 (0.04) | 17.87 (0.04) |

**Supplementary Table 3: Log of the spectroscopic observations of SN2006jc.** Basic information on the spectra presented in this paper. The spectra have been obtained using the 3.58m Telescopio Nazionale Galileo, the 4.2m William Herschel Telescope and 2.56m Nordic Optical Telescope in La Palma; the 1.82m Copernico Telescope of Mt. Ekar; the 1.22m Galileo Telescope in Asiago-Pennar (Italy), the 1.93m Telescope of the Observatoire de Haute-Provence (St. Michel l' Observatoire, France) and the 2.16m NAOC Telescope of the Xinglong Observatory (China).

| Date | JD | Phase from discovery | Instrument | Grism | Range (A) | Resolution (A) |
|---|---|---|---|---|---|---|
| Oct 12 | 2,454,020.73 | +2.46 | TNG+DOLORES | LR-B + LR-R | 3,350-10,000 | |
| Oct 16 | 2,454,024.69 | +6.42 | Ek1.82m+AFOSC | Gm4+Gm2 | 3,490-9,960 | 23-35 |
| Oct 16 | 2,454,024.73 | +6.46 | WHT+ISIS | R600B+R316R | 3,580-5,140, 5,440-8,400 | 2.7+5.0 |
| Oct 16 | 2,454,025.38 | +7.11 | NAOC2.16m+BFOSC | Gm4 | 3,800-9,000 | 15 |
| Oct 18 | 2,454,026.75 | +8.48 | WHT+ISIS | R600B+R316R | 3,560-5,140, 5,440-8,120 | 2.7+5.0 |
| Oct 23 | 2,454,031.74 | +13.47 | NOT+ALFOSC | Gm4 | 3,220-9,130 | 14 |
| Oct 24 | 2,454,032.65 | +14.38 | TNG+DOLORES | LR-B + LR-R | 3,360-10,000 | 13+13 |
| Oct 25 | 2,454,033.65 | +15.38 | OHP+CARELEC | 300 Tr/mm | 3,690-7,320 | 8.5 |
| Oct 26 | 2,454,034.67 | +16.40 | OHP+CARELEC | 300 Tr/mm | 3,690-7,320 | 8.5 |
| Oct 27 | 2,454,035.63 | +17.36 | Ek1.82m+AFOSC | Gm4+Gm2 | 3,490-9,700 | 23-35 |
| Oct 27 | 2,454,035.70 | +17.43 | As1.22m+B&C | 300 Tr/mm | 3,480-7,360 | 25 |
| Oct 28 | 2,454,036.67 | +18.40 | Ek1.82m+AFOSC | Gm4 | 3,490-7,800 | 23 |
| Oct 30 | 2,454,038.69 | +20.42 | Ek1.82m+AFOSC | Gm4 | 3,490-7,800 | 23 |
| Nov 10 | 2,454,049.63 | +31.36 | NOT+ALFOSC | Gm4 | 3,280-9,120 | 14 |
| Nov 14 | 2,454,053.64 | +35.37 | Ek1.82m+AFOSC | Gm4 | 3,500-7,820 | 23 |
| Nov 27 | 2,454,066.56 | +48.29 | Ek1.82m+AFOSC | Gm4+Gm2 | 3,490-10,000 | 23-35 |
| Dec 12 | 2,454,082.37 | +64.10 | NAOC2.16m+BFOSC | Gm4 | 3,890-7,300 | 15 |
| Dec 14 | 2,454,083.58 | +65.31 | TNG+DOLORES | LR-B + LR-R | 3,380-9,600 | 13+13 |

**Supplementary Table 4: Outburst magnitudes, luminosities and masses of nearby LBV candidates and comparison with UGC4904-V1.**

Note that SN2002kg (NGC2403-V37) was not a "giant outburst" but was part of a typical S-doradus phase shown by many stars in the LBV S-Doradus instability strip. The change in visual magnitude $\Delta M$ was due to a $T_{eff}$ variations at constant luminosity, rather than a giant outburst accompanied by a genuine increase on $M_{bol}$.

| Object | $T_{eff}$ | Log L | $M_{max}$ | $\Delta M$ | $M_{bol,max}$ | Initial Mass ($M_\odot$) | Ref. |
|---|---|---|---|---|---|---|---|
| η Carinae | 27,000 | 6.7 | -14.5 | 4-6 | -14 | 150±30 | 2 |
| SN 1961V | | | -17 | ~6 | -17 | 150-200 | 12 |
| SN 1954J | | | -11 | 4 | ≤ -11.6 | 25-120 | 12 |
| HD 5980 | 35,000 | 6.5 | -10.5 | ~ 3 | -13 | 120±25 | 68 |
| NGC2363-V1 | 12,000 | 6.4 | -10.4 | > 4 | -11.3 | 100±20 | 69 |
| SN 1997bs | | | -13.8 | 5.7 | | | 70 |
| SN 2000ch | | | -12.7 | | | | 71 |
| SN 2003gm | | | -13.7 | 5.4 | | | 30 |
| P Cygni | 2,000 | 5.9 | -10.2 | ~ 2-3 | ~ -11 | 60 | 72 |
| SN 2002kg | 16,000 | 5.8 | -10.2 | ~ 2 | -10.5 | 40 | 12 |
| UGC4904-V1 | | | -14.1 | > 2.15 | | | this paper |

References used only in the Supplementary information: